\documentclass[times,twocolumn,tighten]{aastex6}

\usepackage{graphicx}
\usepackage{amsmath}
\usepackage{hyperref}

\DeclareGraphicsExtensions{.pdf}

\shorttitle{Braking Index of PSR~J1208$-$6238}
\shortauthors{\sc Clark et al.}

\begin{document}

\title{The Braking Index of a Radio-quiet Gamma-ray Pulsar}

\author{
    C.~J.~Clark\altaffilmark{1,2,3},
    H.~J.~Pletsch\altaffilmark{1,2},
    J.~Wu\altaffilmark{4},
    L.~Guillemot\altaffilmark{5,6,4},
    F.~Camilo\altaffilmark{7},
    T.~J.~Johnson\altaffilmark{8},
    M.~Kerr\altaffilmark{9},
    B.~Allen\altaffilmark{1,10,2},
    C.~Aulbert\altaffilmark{1,2},
    C.~Beer\altaffilmark{1,2},
    O.~Bock\altaffilmark{1,2},
    A.~Cu\'ellar\altaffilmark{1,2},
    H.~B.~Eggenstein\altaffilmark{1,2},
    H.~Fehrmann\altaffilmark{1,2},
    M.~Kramer\altaffilmark{4,11,12},
    B.~Machenschalk\altaffilmark{1,2},
    and L.~Nieder\altaffilmark{1,2}
  }
  \altaffiltext{1}{Albert-Einstein-Institut, Max-Planck-Institut f\"ur Gravitationsphysik, D-30167
    Hannover, Germany}
  \altaffiltext{2}{Leibniz Universit\"at Hannover, D-30167 Hannover, Germany}
  \altaffiltext{3}{email: colin.clark@aei.mpg.de}
  \altaffiltext{4}{Max-Planck-Institut f\"ur Radioastronomie, Auf dem H\"ugel 69, D-53121 Bonn,
    Germany}
  \altaffiltext{5}{Laboratoire de Physique et Chimie de l'Environnement et de l'Espace --
    Universit\'e d'Orl\'eans / CNRS, F-45071 Orl\'eans Cedex 02, France}
  \altaffiltext{6}{Station de radioastronomie de Nan\c{c}ay, Observatoire de Paris, CNRS/INSU,
    F-18330 Nan\c{c}ay, France}
  \altaffiltext{7}{SKA South Africa, Pinelands, 7405, South Africa.}
  \altaffiltext{8}{College of Science, George Mason University, Fairfax, VA 22030, resident at Naval
    Research Laboratory, Washington, DC 20375, USA}
  \altaffiltext{9}{CSIRO Astronomy and Space Science, Australia Telescope National Facility, Epping
    NSW 1710, Australia}
  \altaffiltext{10}{Department of Physics, University of Wisconsin-Milwaukee, P.O. Box 413,
    Milwaukee, WI 53201, USA}
  \altaffiltext{11}{Jodrell Bank Centre for Astrophysics, School of Physics and Astronomy, The
    University of Manchester, M13 9PL, UK}
  \altaffiltext{12}{University of Manchester, Manchester, M13 9PL, UK}
\begin{abstract}
  \noindent
  We report the discovery and timing measurements of PSR~J1208$-$6238, a young and highly magnetized
  gamma-ray pulsar, with a spin period of $440$~ms. The pulsar was discovered in gamma-ray photon
  data from the \textit{Fermi} Large Area Telescope (LAT) during a blind-search survey of
  unidentified LAT sources, running on the distributed volunteer computing system
  \textit{Einstein@Home}. No radio pulsations were detected in dedicated follow-up searches with the
  Parkes radio telescope, with a flux density upper limit at 1369~MHz of $30\,\mu$Jy. By timing this
  pulsar's gamma-ray pulsations, we measure its braking index over five years of LAT observations to
  be $n = 2.598 \pm 0.001 \pm 0.1$, where the first uncertainty is statistical and the second
  estimates the bias due to timing noise. Assuming its braking index has been similar since birth,
  the pulsar has an estimated age of around 2,700~yr, making it the youngest pulsar to be found in a
  blind search of gamma-ray data and the youngest known radio-quiet gamma-ray pulsar. Despite its
  young age the pulsar is not associated with any known supernova remnant or pulsar wind nebula. The
  pulsar's inferred dipolar surface magnetic field strength is $3.8\times10^{13}$~G, almost $90\%$ of the
  quantum-critical level. We investigate some potential physical causes of the braking index
  deviating from the simple dipole model but find that LAT data covering a longer time interval will
  be necessary to distinguish between these.
\end{abstract}


\section{Introduction}\label{s:intro}
The physical mechanisms by which pulsars radiate rotational energy are as yet unclear. The
dominant process can be inferred by measuring a pulsar's \textit{braking index}, $n$, the index
of a power law relating the pulsar's spin frequency, $f$, to its spin-down rate, $\dot{f}$, via
\begin{equation}
  \dot{f} \propto - f^n\,.
  \label{e:spin_down_model}
\end{equation}
For example, the simple model of a pulsar as a spinning magnetic dipole predicts $n = 3$
\citep{Ostriker1969}.

A pulsar's braking index can be calculated from measurements of the first two time-derivatives of
its spin frequency,
\begin{equation}
  n = \frac{f \ddot{f}}{\dot{f}^2}\,.
  \label{e:braking_index}
\end{equation}
However, most young pulsars exhibit unpredictable fluctuations in their spin frequency on top of
their long-term spin-down behavior \citep{Hobbs2010+TimingNoise} known as \textit{timing
noise}. Observations spanning long time intervals are required to discriminate the overall braking
behavior in these fluctuations.

Braking indices can therefore only be measured for the youngest or most highly magnetized pulsars,
whose long-term braking is still large enough to dominate their spin-down variation. Reliable
measurements of braking indices have been possible for just nine pulsars \citep[and references
  therein]{Archibald2016+J1640,Marshall2016+B0540,Lyne2015+Crab}. All of these braking indices
deviate significantly from $n=3$, with only PSR~J1640$-$4631 having $n>3$
\citep{Archibald2016+J1640}.

One pulsar with a measurable braking index, PSR~J1119$-$6127 \citep{Camilo2000+2HighBPSRs}, is
particularly unusual. Its emission properties (e.g. radio pulsations, exponentially cutoff power-law
gamma-ray spectrum) were, until recently, typical for a ``normal'' rotationally powered pulsar
\citep{Parent2011+J1119}, despite its almost magnetar-level magnetic field
($4.1\times10^{13}$~G). The recent magnetar-like outburst from PSR~J1119$-$6127
\citep{Archibald2016+J1119Flare,Gogus2016+J1119}, and similar events from PSR~J1846$-$0258
\citep{Gavriil2008+J1846Magnetar}, therefore offer insights into the connection between magnetars
and rotationally-powered pulsars.

The Large Area Telescope \citep[LAT; ][]{generalfermilatref} on board the \textit{Fermi Gamma-ray
Space Telescope} has proven to be a valuable instrument for the study of young pulsars
\citep{Grenier2015+Goldmine}. The LAT's $8$ years of almost continuous coverage of the
entire gamma-ray sky has led to the detection of over 200 gamma-ray
pulsars\footnote{\url{http://tinyurl.com/fermipulsars}}. This long observation span enables precise
timing analyses of gamma-ray pulsars \citep[e.g.][]{Ray2011,Kerr2015+FermiTiming,Pletsch2015+J2339}
immediately after their detection.

Blind searches in LAT data have led to the discovery of a sizeable population of young, radio-quiet
gamma-ray pulsars
\citep[e.g.][]{Abdo2009+Blind16,SazParkinson2010+8BSPs,Pletsch+2012-9pulsars}.
However, until now, none of these have had measurable braking indices.

In this Letter, we report the discovery of PSR~J1208$-$6238, a very young, highly magnetized
gamma-ray pulsar with similar properties to those of PSR~J1119$-$6127, including a
measurable braking index. The discovery was made during a large-scale blind-search survey that ran
on the distributed volunteer computing system, \textit{Einstein@Home} \citep[][ApJ, accepted; Wu et
  al. 2016, in preparation]{Clark2015+J1906, Clark2016+FGRP4}.

\section{Observations}
\subsection{LAT Data}\label{s:data}
In the \textit{Einstein@Home} survey, we searched for pulsations in gamma-ray photons from
unidentified sources in the \textit{Fermi}-LAT Third Source Catalog \citep[3FGL; ][]{3FGL}. One such
source was 3FGL~J1208.4$-$6239.

We used ``Pass 8'' \citep{Pass8} LAT data consisting of the arrival times of \texttt{SOURCE}-class
photons above $100$~MeV, and weights, $\left\{w_j\right\}$, representing the probability of each
photon having come from the target source \citep{Kerr2011}. The data in which the pulsar was
discovered spanned 2008~August~4 to 2014~October~1 and were produced using internal preliminary
versions of the Pass 8 instrument response functions (IRFs) and background models. We included
photons from within an $8\arcdeg$ region of interest (ROI) around the 3FGL position, with a maximum
zenith angle of $100\arcdeg$, and a maximum cutoff on the LAT's rocking angle of $52\arcdeg$.

After discovering the pulsar, we produced an extended dataset covering observations until
2016~February~16 for use in follow-up spectral (Section \ref{s:off-pulse}) and timing analyses
(Section \ref{s:timing}). The extended dataset was produced using the \texttt{P8R2\_SOURCE\_V6}
IRFs, a lower zenith angle cutoff of $90\arcdeg$, and a larger $15\arcdeg$ ROI.

We calculated probability weights for photons within $5\arcdeg$ of the target source with
\texttt{gtsrcprob} using the results of a binned likelihood spectral analysis performed with
\texttt{pointlike}. Our source model included all 3FGL sources within a radius $5\arcdeg$ larger
than the ROI. The target source's spectrum was modeled with an exponentially cutoff power law. Its
sky position and spectral parameters were free to vary, as were the spectra of 3FGL sources within
$5\arcdeg$, and the normalizations of the background models. 

When producing the extended dataset, the spectral analysis was performed with \texttt{gtlike}, with
the pulsar's position fixed at its preliminary timing position. The Galactic diffuse emission and
isotropic diffuse background were modeled with the \texttt{gll\_iem\_v06.fits}
\citep{Acero2016+Diffuse} and \texttt{iso\_P8R2\_SOURCE\_V6\_v06.txt}
templates\footnote{\url{http://fermi.gsfc.nasa.gov/ssc/data/access/lat/BackgroundModels.html}}
respectively.

For the blind search and subsequent timing analyses we performed minimum weight cutoffs to reduce
the number of photons for computational speed. In the blind search, this cutoff was $w_{\rm
  min}=3.9\%$, leaving $\sum w_j=2762.7$ ``effective'' photons. In the timing analysis, $w_{\rm
  min}=0.7\%$ and $\sum w_j=5284.4$.

\subsection{Discovery in a Blind Search}\label{s:search}
The first stage of the survey employed a ``semicoherent'' search, in which only photons arriving
within $2^{21}$~s ($\approx24$~days) of one another were combined coherently
\citep{Methods2014}. Candidates from this stage were then ``followed up'' in more sensitive
(ultimately fully coherent) stages to increase their significance. While our search model assumed a
constant spin-down rate, the semicoherent stage was sensitive to signals with varying spin-down,
provided $\left|\ddot{f}\right|\lesssim3\times10^{-22}$~Hz~s$^{-2}$.
However, pulsars with ${\left|\ddot{f}\right|\gtrsim5\times10^{-24}}$~Hz~s${^{-2}}$ would be missed by
the coherent follow-up stage, and could only be detected if their signal was strong enough to be
significant in the initial semicoherent stage.

Visual inspection of a candidate signal from 3FGL~J1208.4$-$6239 revealed clear pulsations,
identifying this source as the pulsar now known as PSR~J1208$-$6238. The initial signal showed large
variations in the pulse phase. Including a second frequency derivative of
$3.3\times10^{-22}$~Hz~s$^{-2}$ in the phase model removed the majority of these.

\subsection{Off-pulse Analysis}\label{s:off-pulse}
The integrated pulse profile from our initial timing solution showed significant emission above the
background at all phases, indicative either of flux from the pulsar or from a nearby unmodeled
source.

\begin{figure}
	\centering
	\includegraphics[width=\columnwidth]{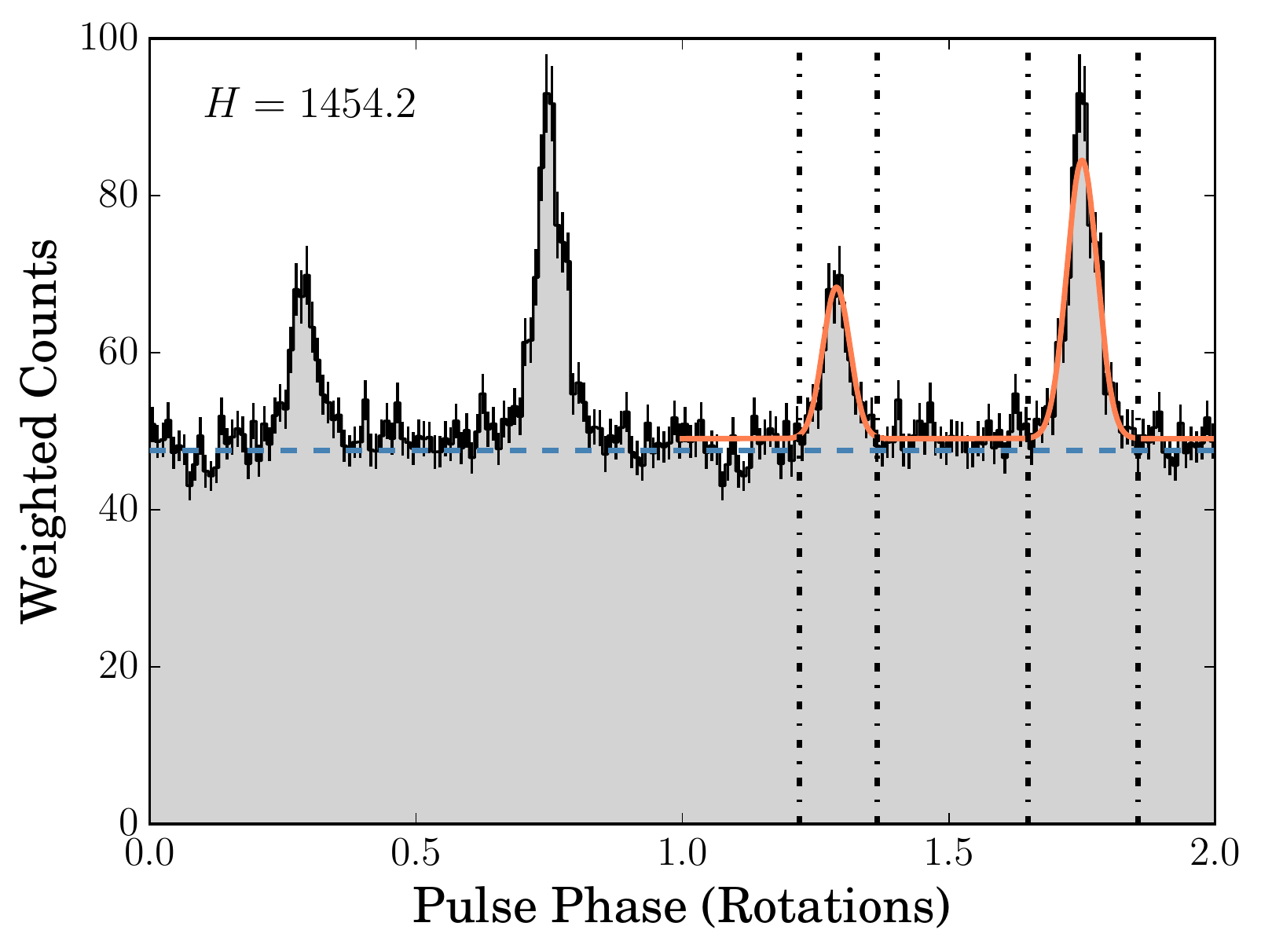}
	\caption{Gamma-ray pulse profile of PSR~J1208$-$6238, using the photon weights described in
    Section \ref{s:off-pulse} and the Taylor series timing solution from Section \ref{s:timing}.
    Two rotations are plotted; the second shows the template pulse profile (orange curve) used in
    the timing analysis (Section \ref{s:timing}). Vertical dash-dotted lines mark the phase ranges
    excluded from the off-pulse analysis (Section~\ref{s:off-pulse}). The dashed blue line shows the
    background level, estimated from the photon probability weights as described in
    \citet{2PC+2013}.}
	\label{f:pulse_profile}
\end{figure}

To investigate this unpulsed flux, we assigned rotational phases to all photons using our initial
timing solution, and performed a spectral analysis of the off-pulse emission (i.e. excluding photons
with phases within the ranges shown in Figure~\ref{f:pulse_profile}). This revealed additional
nearby gamma-ray sources, the closest and most significant of which lies $20\arcmin$ from the
pulsar's timing position, with a spectral index of $\Gamma=2.56\pm0.09$ and $(40\pm13)\%$ of the
energy flux of PSR~J1208$-$6238 above 100~MeV.

The nature of these sources is unclear; without likely counterparts at other wavelengths it is
difficult to tell whether these are point sources or residuals from the Galactic diffuse emission
template. Their steep spectra are consistent with LAT-detected supernova remnants
\citep[SNRs;][]{Acero2016+SNRCat}, but kick velocity requirements make an association with the
pulsar unlikely (see Section~\ref{s:discussion}).

After including these additional sources in our model, re-fitting the ROI and re-calculating the
photon weights for PSR~J1208$-$6238, the pulsation significance increased. These weights were
subsequently used in the timing analyses described in Section \ref{s:timing}, resulting in the pulse
profile shown in Figure \ref{f:pulse_profile}, with a final $H$-test \citep{deJager+1989} value
$1454.2$. The pulsar's spectral properties are given in Table \ref{t:timing_solution}.

\subsection{Radio Observations}\label{s:radio}
On 2016 March 28, 2016 April 14 and 2016 June 18 we observed the pulsar timing position with the
64-m Parkes radio telescope for 2.5\,hr, 4.3\,hr and 1.5\,hr respectively using the H-OH receiver at
a center frequency of 1369\,MHz.  With PDFB4 we recorded 256\,MHz of bandwidth filtered into
512-channel spectra with 256\,$\mu$s sampling (80\,$\mu$s for the third observation). The data were
analyzed using standard pulsar search techniques implemented in \texttt{PRESTO}
\citep{Ransom2001+Thesis}. Folding the data with the gamma-ray ephemeris, we searched over
dispersion measure (DM) from 0--2700~pc~cm$^{-3}$. We found no plausible pulsar candidates.

Recent flux-calibrated observations using the same configuration show a root-mean-square noise level
of 130\,$\mu$Jy hr$^{-1}$ for a 32-bin pulse profile.  With a detection threshold of one or more
phase bins with S/N $>$8, then optimally our longest observation would have detected a source with a
mean flux density $>$17\,$\mu$Jy.  Accounting for scalloping losses due to binning in time and DM,
and for the unknown pulse duty cycle, we estimate an upper limit of 30\,$\mu$Jy, equal to the
radio-quiet threshold defined by \citet{2PC+2013}.

\begin{deluxetable}{lc}
  \tabletypesize{\footnotesize}
  \tablewidth{\columnwidth}
  \tablecaption{\label{t:timing_solution} Properties of PSR~J1208$-$6238}
  \tablecolumns{2}
  \tablehead{ \colhead{Parameter} & \colhead{Value}}
  \startdata
Range of photon data (MJD) & 54682--57434 \\[-0.5ex] 
Reference epoch, $t_{\rm ref}$ (MJD) & $56040$ \\[-0.5ex] 
\cutinhead{Timing parameters\tablenotemark{a}}
R.A. (J2000.0) & $12^{\rm h}\,08^{\rm m}\,13^{\rm s}.96(6)$ \\[-0.5ex] 
Decl. (J2000.0) & $-62\,\arcdeg38\,\arcmin02\farcs3(4)$ \\[-0.5ex] 
Spin frequency, $f$ (Hz) & $2.26968010518(7)$ \\[-0.5ex] 
Spin-down rate, $\dot{f}$ ($10^{-12}$ Hz s$^{-1}$) & $-16.842733(5)$ \\[-0.5ex] 
Braking index, $n$ & $2.598(1)$ \\[-0.5ex] 
$\dot{f}$-increment epoch (MJD) & $55548(23)$ \\[-0.5ex] 
$\dot{f}$-increment, $\Delta \dot{f}$ ($10^{-15}$ Hz s$^{-1}$) & $0.59(9)$ \\[-0.5ex] 
$n$-increment, $\Delta n$ & $-0.10(2)$ \\[-0.5ex] 
\cutinhead{Derived properties\tablenotemark{b}}
Galactic longitude, $l$ (\arcdeg) & $297.99$ \\[-0.5ex] 
Galactic latitude, $b$ (\arcdeg) & $-0.18$ \\[-0.5ex] 
Spin period, $P$ (ms) & $440.59072365(1)$ \\[-0.5ex] 
Period derivative, $\dot{P}$ ($10^{-12}$ s s$^{-1}$) & $3.2695145(9)$ \\[-0.5ex] 
Surface B-field stength, $B_{\rm S}$ ($10^{12}$ G) & $38.4$ \\[-0.5ex] 
Estimated age\tablenotemark{c}, $\tau$ (yr) & $2672$ \\[-0.5ex] 
Spin-down power, $\dot{E}$ ($10^{36}$ erg s$^{-1}$) & $1.5$\\[-0.5ex] 
Maximum distance, $d_{100\%}$ (kpc) & $18.9$\\[-0.5ex] 
Heuristic distance, $d_{\rm h}$ (kpc) & $3.0$\\[-0.5ex] 
\cutinhead{Spectral parameters above 100 MeV\tablenotemark{d}}
Spectral index, $\Gamma$ & $1.73 \pm 0.08 \pm 0.04$ \\[-0.5ex] 
Cutoff energy, $E_{\rm c}$~(GeV) & $4.86 \pm 0.59 \pm 0.70$ \\[-0.5ex] 
Photon flux, $F_{100}$~(cm$^{-2}$~s$^{-1}$) & $(4.41 \pm 0.86 \pm 0.37) \times 10^{-8}$  \\[-0.5ex] 
Energy flux, $G_{100}$~(erg cm$^{-2}$~s$^{-1}$) & $(3.49 \pm 0.44 \pm 0.29) \times 10^{-11}$  \\[-0.5ex] 
\enddata
 \tablecomments{The reported values for $f$ and $\dot{f}$ at the reference time include the effect
   of the earlier $\dot{f}$ increment.}
 \tablenotetext{a}{For timing parameters, we report mean
   values and $1\sigma$ uncertainties on the final digits in brackets from the results of the timing
   fit to Equation~\ref{e:powerlaw_phase_model} described in Section \ref{s:timing}.}
 \tablenotetext{b}{Derived properties are calculated as described in \citet{2PC+2013}. Maximum and heuristic distances are calculated assuming isotropic emission and gamma-ray luminosities of $\dot{E}$ and $\sqrt{10^{33}\dot{E}}$ respectively.}
 \tablenotetext{c}{The estimated age was calculated using the measured braking index.}
 \tablenotetext{d}{The first uncertainty is statistical, the second estimates systematic
   uncertainties in the LAT's effective area, estimated by performing the same spectral analysis with
   rescaled effective areas (see
   \url{http://fermi.gsfc.nasa.gov/ssc/data/analysis/scitools/Aeff_Systematics.html} for details), and in the
   Galactic diffuse emission model, estimated by performing the spectral analysis with the
   normalization of the Galactic diffuse emission rescaled to $\pm6\%$ of the best-fit value.}
 \end{deluxetable}

\section{Timing Analysis}\label{s:timing}
To investigate the braking properties of PSR~J1208$-$6238, we performed a timing analysis, following
the procedure described in \citet{Pletsch2015+J2339} and \citet{Clark2015+J1906}, a modification of
the methods developed by \citet{Ray2011}.

Our phase model consisted of a Taylor series,
\begin{equation}
  \Phi(t)=\sum_{m=0}\frac{f^{(m)}}{(m+1)!}(t-t_{\rm ref})^{m+1}\,,
  \label{e:series_phase_model}
\end{equation}
where $f^{(m)}$ denotes the $m$-th time-derivative of the pulsar's rotational frequency at the
reference epoch, $t_{\rm ref}$, and $t$ is the (sky location-dependent) barycentric time. 

Starting with our initial solution, consisting of the pulsar's sky position, frequency and
first two frequency derivatives ($\dot{f}\equiv f^{(1)}$ and $\ddot{f}\equiv f^{(2)}$), we
phase-folded the photon data and fit a template pulse profile (shown in Figure
\ref{f:pulse_profile}) by maximizing the likelihood.  We used the Bayesian Information Criterion
\citep[BIC;][]{Schwarz1978+BIC} to estimate the appropriate number of components to include in the
template, finding that two wrapped Gaussian peaks were sufficient.

We then sampled the pulsar's spin and positional parameters with an Affine Invariant Monte-Carlo
algorithm \citep{Goodman2010+AIMC,Foreman-Mackey2013+emcee}, using the template pulse profile to
evaluate the likelihood at each point. The process continued iteratively; after each sampling stage
the photons were re-folded with the most likely parameters and the template pulse profile was
updated. Higher frequency derivatives were added to the phase model after each stage, again using
the BIC to find the appropriate number of terms.

The solution that optimized the BIC contained nine frequency derivatives. However, the resulting
phase residuals contain some remaining ``red'' noise. We were unable to remove this effect by
including higher frequency derivatives, up to the 12th derivative.

We can calculate $n$ with Equation (\ref{e:braking_index}) using the measured values of $\dot{f}$
and $\ddot{f}$, finding $n=2.578\pm 0.007$. However, as explained by \citet{Antonopoulou2015+J1119},
truncating the Taylor series at the second derivative means that the measured value of $n$ depends
on the arbitrarily chosen reference epoch.  Making the simple assumption of a constant braking
index leads to a self-consistent, physically motivated phase model, which avoids this problem
\citep{Antonopoulou2015+J1119},
\begin{equation}
    \Phi(t)=\frac{f^2}{\dot{f}(2-n)}\left(\left[1+\frac{\dot{f}}{f}(1-n)(t-t_{\rm
      ref})\right]^{\frac{2-n}{1-n}}-1\right)\,.
  \label{e:powerlaw_phase_model}
\end{equation}
In this model, one only fits for $f$, $\dot{f}$, $n$ and sky location. Faster timing noise variations
on top of this long term spin-down are left unmodelled.

\begin{figure}
	\centering
	\includegraphics[width=\columnwidth]{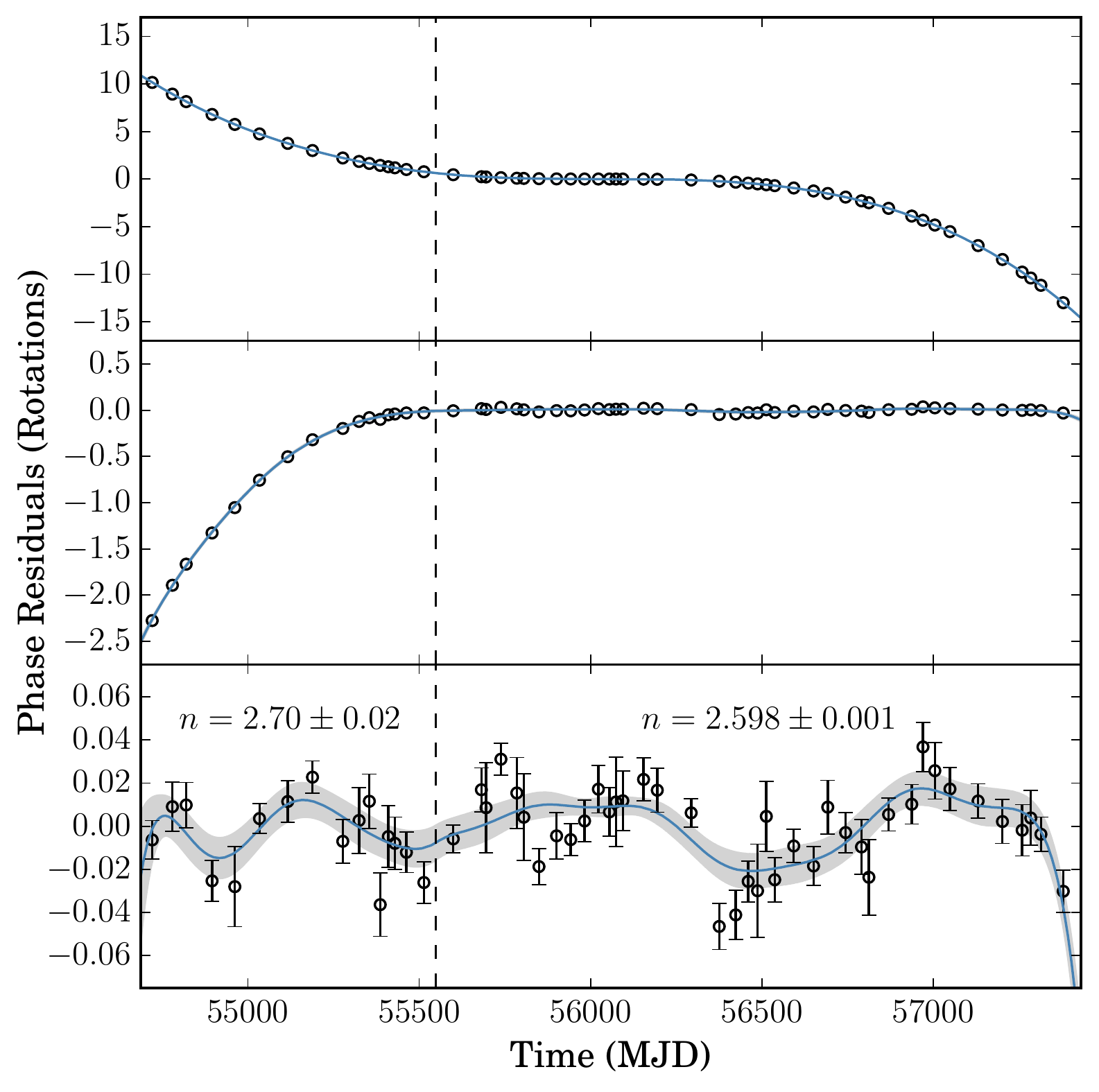}
	\caption{Measured phase residuals from our timing models. The blue lines and gray shaded regions
    represent the best-fitting Taylor series phase model and 1$\sigma$ uncertainties. The ``TOAs''
    shown here were not used to perform the fit, which was an entirely unbinned likelihood
    maximization, but are included here to illustrate the validity of the phase models. Upper panel:
    residuals between the Taylor series and a pure dipole-braking model with $n=3$. The decreasing
    cubic curve is characteristic of a over-estimated braking index. Middle panel:
    residuals between the Taylor series model and a best-fitting constant braking index model. A
    significant deviation is evident in the early mission. Lower panel: as above, but with
    increments in the braking index and spin-down rate occurring at the time marked by the dashed
    vertical line.}
	\label{f:phase_residuals}
\end{figure}

Carrying out this fit for PSR~J1208$-$6238, we find that a constant braking index fits well in the
last five years of the data set, but cannot account for the pulsar's spin-down over the entire data
set, as is clear from Figure \ref{f:phase_residuals}. Instead, we find that increments in the
pulsar's braking index and spin-down rate are required. The results of this fit, including these
increments, are given in Table \ref{t:timing_solution}. We cannot tell whether these were sudden or
gradual changes as it takes many months for an offset in either of these parameters to cause a
detectable phase offset. We note that \citet{Archibald2015+J1846} required a similar change in
$\dot{f}$ and $n$ in their timing model for PSR~J1846$-$0258. We find no evidence for changes in the
pulsar's pulse profile or gamma-ray flux \citep[like those of PSR~J2021$+$4026,][]{Allafort2013+J2021+4026} associated with the $\dot{f}$ increment.

\section{Discussion}\label{s:discussion}
PSR~J1208$-$6238 is the first radio-quiet gamma-ray pulsar with a measured braking index, and only
the tenth pulsar of any kind with such a measurement. Its estimated age, assuming its braking index
has been constant since birth, is around 2,700~yr, making it the youngest known radio-quiet
gamma-ray pulsar.

PSR~J1208$-$6238 shares many similarities (e.g. age, spin-down power) with PSR~J1119$-$6127, an
otherwise unique gamma-ray pulsar. They are both highly magnetized, with estimated dipolar surface
magnetic field strengths (accurate to an order-of-magnitude), $B_{\rm
  S}\sim3.2\times10^{19}(-\dot{f}f^{-3})^{1/2}$, of $3.8\times10^{13}$~G (PSR~J1208$-$6238) and
$4.1\times10^{13}$~G (PSR~J1119$-$6127), close to the quantum-critical limit $B_{\rm
  cr}=4.4\times10^{13}$~G \citep{Baring1998+HighBRadioQuiet}. Such a high magnetic field could
affect a pulsar's high-energy emission, however both pulsars have gamma-ray spectra that are typical
of young pulsars \citep{2PC+2013}. Given their similar magnetic fields, one may expect
PSR~J1208$-$6238 to exhibit similar magnetar-like activity as was recently seen from
PSR~J1119$-$6127 \citep{Archibald2016+J1119Flare,Gogus2016+J1119}. However, no X-ray emission has
been detected from PSR~J1208$-$6238 \citep{Swift+LATSurvey}.

While radio pulsations have been detected from transient magnetars in outburst states
\citep[e.g.][]{Camilo2006+J1810}, radio pulsations can be suppressed by strong magnetic fields
\citep{Baring1998+HighBRadioQuiet,Camilo2000+2HighBPSRs}.  However, perhaps a more likely
explanation for this pulsar's radio-quietness is that its radio beam simply does not cross our
line-of-sight.

To investigate this possibility we modeled the gamma-ray emission geometry using the methods of
\citet{Johnson2014+LCModelling}, with outer gap (OG) and two-pole caustic (TPC) models described
therein. Our simulations used\footnote{While these values
differ from those in Table~\ref{t:timing_solution} 
the gamma-ray profile dependence on these is weak at these values so our conclusions will not be
affected.} $P=100$~ms and $\dot{P}=10^{-15}$~s~s$^{-1}$. We used steps in gap widths of 1\% the
polar cap opening angle. Model profiles were fit using 
weighted photon counts and the $\chi^2$ statistic, rather than the Poisson likelihood used in
\citet{Johnson2014+LCModelling}. The TPC model is slightly favored, although not significantly, with
the angle between the line-of-sight and magnetic angle, $\beta=2\substack{+1\\-1}\,\arcdeg$,
predicting visible radio pulsations. The OG fit gives $\beta=28\substack{+8\\-6}\,\arcdeg$, with
no visible radio emission. We caution that performing these fits without constraints on
the radio emission can lead to large systematic errors \citep{Pierbattista2015+LCM}, and note that
neither model perfectly reproduces the observed peak separation.

We also note that, due to its unknown distance, our flux density upper limit for PSR~J1208$-$6238
does not strongly constrain its radio luminosity. Its radio pulsations could simply be too faint to be
detected.

Despite the pulsar's very young inferred age, no positionally-associated SNR or pulsar wind nebula
has been detected by radio imaging \citep{Murphy2007+MGPS2}, X-ray observations
\citep{Swift+LATSurvey,Hwang1994+ROSATSNRs}, or in TeV emission \citep{TeVCat}.  There is, however,
a luminous \ion{H}{2} region \citep[IRAS~12073$-$6233; ][]{MartinHernandez2003+HII} less than
$0.5\arcdeg$ away from PSR~J1208$-$6238, whose radio emission could mask a faint SNR near the
pulsar. A dedicated search for an associated SNR may be required to clarify this. The two nearest
known SNRs (G298.6$-$0.0 and G298.5$-$0.3) are more than half a degree away from the pulsar
\citep{Green+SNRCat}. Due to the pulsar's young age, an association with either of these SNRs
implies an unrealistically high kick velocity, approximately $3000\,(d/1\,{\rm kpc})$~km~s$^{-1}$
for pulsar distance $3\,{\rm kpc}\lesssim d<19\,{\rm kpc}$ (see
Table~\ref{t:timing_solution}). A bias on the pulsar's timing position due to timing noise
\citep{Kerr2015+FermiTiming} is insufficient to account for the positional offsets to the nearest
SNRs.

The braking index measurement is also sensitive to the pulsar's intrinsic timing noise. The
uncertainty on $n$ of $\pm10^{-3}$ quoted in Table~\ref{t:timing_solution} is statistical only;
timing noise will lead to an additional unknown bias. Assuming that the observed change in $n$
around MJD~$55600$ was due to timing noise, we take this increment, $\Delta n=0.1$, as an estimate
of this systematic uncertainty.

The larger braking index in the earlier data could alternatively be caused by the pulsar relaxing
from a glitch occurring before the start of the \textit{Fermi} mission, of the kind observed from
PSR~J1119$-$6127 \citep[e.g.,][]{Antonopoulou2015+J1119}. However, a timing model featuring an
exponentially decaying frequency term in the early mission is disfavored by the BIC.

Nevertheless, our timing measurements constrain the braking index to be below the $n=3$ predicted
by a dipole braking model, as evident from the large cubic residuals shown in Figure
\ref{f:phase_residuals}. A reduced braking index can be explained by the pulsar's physical
properties varying over time. Differentiation of the dipole braking model (assuming constant moment
of inertia) gives \citep{Lyne2015+Crab},
\begin{equation}
  n = 3+2\frac{f}{\dot{f}}\left[\frac{\dot{\mu}}{\mu}+\frac{\dot{\alpha}}{\tan\alpha}\right]\,,
\end{equation}
where $\mu$ is the magnetic dipole moment, inclined at an angle $\alpha$ to the spin axis. The
measured braking index can be explained by fractional changes in these parameters of
$\sim5\times10^{-5}$~yr$^{-1}$.

A low braking index can be explained by a growing magnetic field \citep{Blandford1988}. This could
be caused by the pulsar's initial magnetic field being ``buried'' by matter accretion shortly after
birth, and gradually growing back to its original strength \citep{Ho2015+MagFieldGrowth}. This
implies a braking index that evolves back to $3$ over a timescale of $\sim10^{5}$~yr. This evolution
of $n$ is undetectable with current observation lengths.

A varying $\alpha$ leads to evolution of the observed pulse profile. This has been observed
from the Crab pulsar, where a measured increase in the (angular) peak separation of $\approx
6\times10^{-3}$~$\arcdeg$~yr$^{-1}$ implies a similar magnitude for $\dot{\alpha}$
\citep{Lyne2013+CrabAlphaDot}. The observed $\dot{\alpha}$, which could under certain conditions be
attributed to precession \citep{Arzamasskiy2015+precession}, is sufficient to explain the Crab's
low braking index. 

Fits to TPC and OG models described above estimate $\alpha=81\substack{+1 \\ -1}\,\arcdeg$ and
$\alpha=57\substack{+8 \\ -5}\,\arcdeg$ respectively for PSR~J1208$-$6238. For the variation in
$\alpha$ to account for the reduced braking index requires
$\dot{\alpha}\approx2\times10^{-2}$~$\arcdeg$~yr$^{-1}$ (TPC) or $\dot{\alpha}\approx
4\times10^{-3}$~$\arcdeg$~yr$^{-1}$ (OG). For either model, the expected evolution of the gamma-ray
pulse profile caused by the required $\dot{\alpha}$ is too small to be measured with the current
data.

Low braking indices can also be explained by a portion of the total spin-down torque being applied
by a process with a different braking index, e.g. angular momentum being lost to an outflowing
particle wind \citep[$n=1$; ][]{Harding1999+Wind}, or propeller torque from an in-falling
disk \citep[$n = -1$; ][]{Menou2001+Disk}. The fraction, $\epsilon$, of the total spin-down torque
that a process with a braking index of $n_2$ must account for to explain an observed index $n$
is\footnote{ \citet{Lyne2015+Crab} define $\epsilon$ as the ratio of wind-induced torque to
  dipole-induced torque, rather than wind-induced torque to total spin-down torque. This definition
  may have been used in error by \citet{Archibald2015+J1846,Archibald2016+J1640}, although their
  conclusions are unchanged.}
\begin{equation}
  \epsilon\approx\frac{3-n}{3-n_2}\,.
  \label{e:epsilon}
\end{equation}
Under these models, as the pulsar spins down
$\epsilon$ will increase (provided $n_2<3$), leading to a time-varying braking index, where
\begin{equation}
  \dot{n}\approx\frac{\dot{f}}{f}\epsilon\left(n_2-3\right)\left(n_2-n\right)\,.
  \label{e:ndot}
\end{equation}
For the above wind (disk) model, we find $\epsilon \approx 20\%$ ($10\%$) and $\dot{n} \approx
-1.5\times10^{-4}$~yr$^{-1}$ ($-3.4\times10^{-4}$~yr$^{-1}$). The braking index
variations are not currently measurable, but may become so with future LAT data.\\

\acknowledgements We gratefully acknowledge the support of all volunteers who have donated CPU
cycles to \textit{Einstein@Home}. We are especially grateful to James Drews of UW-Madison, WI, USA and
University of Houston, IT High Performance Computing, TX, USA, whose computers discovered PSR~J1208$-$6238.

This work was supported by the Max-Planck-Gesell\-schaft~(MPG), by the Deutsche
Forschungsgemeinschaft~(DFG) through an Emmy Noether research grant PL~710/1-1 (PI:
Holger~J.~Pletsch), and by NSF award 1104902.

The Parkes radio telescope is part of the Australia Telescope, which is funded by the Commonwealth
Government for operation as a National Facility managed by CSIRO.

The \textit{Fermi}-LAT Collaboration acknowledges support for LAT development, operation and data analysis from NASA and DOE (United States), CEA/Irfu and IN2P3/CNRS (France), ASI and INFN (Italy), MEXT, KEK, and JAXA (Japan), and the K.A.~Wallenberg Foundation, the Swedish Research Council and the National Space Board (Sweden). Science analysis support in the operations phase from INAF (Italy) and CNES (France) is also gratefully acknowledged.


\begin{thebibliography}{}
\expandafter\ifx\csname natexlab\endcsname\relax\def\natexlab#1{#1}\fi

\bibitem[{{Abdo} {et~al.}(2009){Abdo}, {Ackermann}, {Ajello},
  {et~al.}}]{Abdo2009+Blind16}
{Abdo}, A.~A., {Ackermann}, M., {Ajello}, M., {et~al.} 2009, Science, 325, 840

\bibitem[{{Abdo} {et~al.}(2013){Abdo}, {Ajello}, {Allafort},
  {et~al.}}]{2PC+2013}
{Abdo}, A.~A., {Ajello}, M., {Allafort}, A., {et~al.} 2013, \apjs, 208, 17

\bibitem[{{Acero} {et~al.}(2016{\natexlab{a}}){Acero}, {Ackermann}, {Ajello},
  {Albert}, {Baldini}, {Ballet}, {et~al.}}]{Acero2016+Diffuse}
{Acero}, F., {Ackermann}, M., {Ajello}, M., {et~al.} 2016{\natexlab{a}}, \apjs,
  223, 26

\bibitem[{{Acero} {et~al.}(2016{\natexlab{b}}){Acero}, {Ackermann}, {Ajello},
  {Baldini}, {Ballet}, {Barbiellini}, {et~al.}}]{Acero2016+SNRCat}
---. 2016{\natexlab{b}}, \apjs, 224, 8

\bibitem[{{Acero} {et~al.}(2015){Acero}, {Ackermann}, {Ajello},
  {et~al.}}]{3FGL}
---. 2015, \apjs, 218, 23

\bibitem[{{Allafort} {et~al.}(2013){Allafort}, {Baldini}, {Ballet},
  {Barbiellini}, {Baring}, {Bastieri}, {Bellazzini},
  {et~al.}}]{Allafort2013+J2021+4026}
{Allafort}, A., {Baldini}, L., {Ballet}, J., {et~al.} 2013, \apjl, 777, L2

\bibitem[{{Antonopoulou} {et~al.}(2015){Antonopoulou}, {Weltevrede},
  {Espinoza}, {Watts}, {Johnston}, {Shannon}, \&
  {Kerr}}]{Antonopoulou2015+J1119}
{Antonopoulou}, D., {Weltevrede}, P., {Espinoza}, C.~M., {et~al.} 2015, \mnras,
  447, 3924

\bibitem[{{Archibald} {et~al.}(2015){Archibald}, {Kaspi}, {Beardmore},
  {Gehrels}, \& {Kennea}}]{Archibald2015+J1846}
{Archibald}, R.~F., {Kaspi}, V.~M., {Beardmore}, A.~P., {Gehrels}, N., \&
  {Kennea}, J.~A. 2015, \apj, 810, 67

\bibitem[{{Archibald} {et~al.}(2016{\natexlab{a}}){Archibald}, {Kaspi},
  {Tendulkar}, \& {Scholz}}]{Archibald2016+J1119Flare}
{Archibald}, R.~F., {Kaspi}, V.~M., {Tendulkar}, S.~P., \& {Scholz}, P.
  2016{\natexlab{a}}, \apjl, 829, L21

\bibitem[{{Archibald} {et~al.}(2016{\natexlab{b}}){Archibald}, {Gotthelf},
  {Ferdman}, {Kaspi}, {Guillot}, {Harrison}, {Keane}, {Pivovaroff}, {Stern},
  {Tendulkar}, \& {Tomsick}}]{Archibald2016+J1640}
{Archibald}, R.~F., {Gotthelf}, E.~V., {Ferdman}, R.~D., {et~al.}
  2016{\natexlab{b}}, \apjl, 819, L16

\bibitem[{{Arzamasskiy} {et~al.}(2015){Arzamasskiy}, {Philippov}, \&
  {Tchekhovskoy}}]{Arzamasskiy2015+precession}
{Arzamasskiy}, L., {Philippov}, A., \& {Tchekhovskoy}, A. 2015, \mnras, 453,
  3540

\bibitem[{{Atwood} {et~al.}(2012){Atwood}, {Albert}, {Baldini},
  {et~al.}}]{Pass8}
{Atwood}, W., {Albert}, A., {Baldini}, L., {et~al.} 2012, in Proceedings of the
  4th Fermi Symposium, ed. T.~J. {Brandt}, N.~{Omodei}, \& C.~{Wilson-Hodge},
  eConf C121028, 8, arXiv:1303.3514

\bibitem[{Atwood {et~al.}(2009)Atwood, Abdo, Ackermann,
  {et~al.}}]{generalfermilatref}
Atwood, W.~B., Abdo, A.~A., Ackermann, M., {et~al.} 2009, \apj, 697, 1071

\bibitem[{{Baring} \& {Harding}(1998)}]{Baring1998+HighBRadioQuiet}
{Baring}, M.~G., \& {Harding}, A.~K. 1998, \apjl, 507, L55

\bibitem[{{Blandford} \& {Romani}(1988)}]{Blandford1988}
{Blandford}, R.~D., \& {Romani}, R.~W. 1988, \mnras, 234, 57P

\bibitem[{{Camilo} {et~al.}(2000){Camilo}, {Kaspi}, {Lyne}, {Manchester},
  {Bell}, {D'Amico}, {McKay}, \& {Crawford}}]{Camilo2000+2HighBPSRs}
{Camilo}, F., {Kaspi}, V.~M., {Lyne}, A.~G., {et~al.} 2000, \apj, 541, 367

\bibitem[{{Camilo} {et~al.}(2006){Camilo}, {Ransom}, {Halpern}, {Reynolds},
  {Helfand}, {Zimmerman}, \& {Sarkissian}}]{Camilo2006+J1810}
{Camilo}, F., {Ransom}, S.~M., {Halpern}, J.~P., {et~al.} 2006, \nat, 442, 892

\bibitem[{{Clark} {et~al.}(2015){Clark}, {Pletsch}, {Wu},
  {et~al.}}]{Clark2015+J1906}
{Clark}, C.~J., {Pletsch}, H.~J., {Wu}, J., {et~al.} 2015, \apjl, 809, L2

\bibitem[{{Clark} {et~al.}(2016){Clark}, {Wu}, {Pletsch}, {Guillemot}, {Allen},
  {Aulbert}, {Beer}, {Bock}, {Cu{\'e}llar}, {Eggenstein}, {Fehrmann}, {Kramer},
  {Machenschalk}, \& {Nieder}}]{Clark2016+FGRP4}
{Clark}, C.~J., {Wu}, J., {Pletsch}, H.~J., {et~al.} 2016, ArXiv e-prints,
  arXiv:1611.01015

\bibitem[{{de Jager} {et~al.}(1989){de Jager}, {Raubenheimer}, \&
  {Swanepoel}}]{deJager+1989}
{de Jager}, O.~C., {Raubenheimer}, B.~C., \& {Swanepoel}, J.~W.~H. 1989, \aap,
  221, 180

\bibitem[{{Foreman-Mackey} {et~al.}(2013){Foreman-Mackey}, {Hogg}, {Lang}, \&
  {Goodman}}]{Foreman-Mackey2013+emcee}
{Foreman-Mackey}, D., {Hogg}, D.~W., {Lang}, D., \& {Goodman}, J. 2013, \pasp,
  125, 306

\bibitem[{{Gavriil} {et~al.}(2008){Gavriil}, {Gonzalez}, {Gotthelf}, {Kaspi},
  {Livingstone}, \& {Woods}}]{Gavriil2008+J1846Magnetar}
{Gavriil}, F.~P., {Gonzalez}, M.~E., {Gotthelf}, E.~V., {et~al.} 2008, Science,
  319, 1802

\bibitem[{{Goodman} \& {Weare}(2010)}]{Goodman2010+AIMC}
{Goodman}, J., \& {Weare}, J. 2010, "Comm. App. Math. Comp. Sci.", 5, 65

\bibitem[{{G{\"o}{\u g}{\"u}{\c s}} {et~al.}(2016){G{\"o}{\u g}{\"u}{\c s}},
  {Lin}, {Kaneko}, {Kouveliotou}, {Watts}, {Chakraborty}, {Alpar},
  {Huppenkothen}, {Roberts}, {Younes}, \& {van der Horst}}]{Gogus2016+J1119}
{G{\"o}{\u g}{\"u}{\c s}}, E., {Lin}, L., {Kaneko}, Y., {et~al.} 2016, \apjl,
  829, L25

\bibitem[{{Green}(2014)}]{Green+SNRCat}
{Green}, D.~A. 2014, BASI, 42, 47

\bibitem[{{Grenier} \& {Harding}(2015)}]{Grenier2015+Goldmine}
{Grenier}, I.~A., \& {Harding}, A.~K. 2015, Comptes Rendus Physique, 16, 641

\bibitem[{{Harding} {et~al.}(1999){Harding}, {Contopoulos}, \&
  {Kazanas}}]{Harding1999+Wind}
{Harding}, A.~K., {Contopoulos}, I., \& {Kazanas}, D. 1999, \apjl, 525, L125

\bibitem[{{Ho}(2015)}]{Ho2015+MagFieldGrowth}
{Ho}, W.~C.~G. 2015, \mnras, 452, 845

\bibitem[{{Hobbs} {et~al.}(2010){Hobbs}, {Lyne}, \&
  {Kramer}}]{Hobbs2010+TimingNoise}
{Hobbs}, G., {Lyne}, A.~G., \& {Kramer}, M. 2010, \mnras, 402, 1027

\bibitem[{{Hwang} \& {Markert}(1994)}]{Hwang1994+ROSATSNRs}
{Hwang}, U., \& {Markert}, T.~H. 1994, \apj, 431, 819

\bibitem[{{Johnson} {et~al.}(2014){Johnson}, {Venter}, {Harding}, {Guillemot},
  {Smith}, {Kramer}, {{\c C}elik}, {den Hartog}, {Ferrara}, {Hou}, {Lande}, \&
  {Ray}}]{Johnson2014+LCModelling}
{Johnson}, T.~J., {Venter}, C., {Harding}, A.~K., {et~al.} 2014, \apjs, 213, 6

\bibitem[{{Kerr}(2011)}]{Kerr2011}
{Kerr}, M. 2011, \apj, 732, 38

\bibitem[{{Kerr} {et~al.}(2015){Kerr}, {Ray}, {Johnston}, {Shannon}, \&
  {Camilo}}]{Kerr2015+FermiTiming}
{Kerr}, M., {Ray}, P.~S., {Johnston}, S., {Shannon}, R.~M., \& {Camilo}, F.
  2015, \apj, 814, 128

\bibitem[{{Lyne} {et~al.}(2013){Lyne}, {Graham-Smith}, {Weltevrede}, {Jordan},
  {Stappers}, {Bassa}, \& {Kramer}}]{Lyne2013+CrabAlphaDot}
{Lyne}, A., {Graham-Smith}, F., {Weltevrede}, P., {et~al.} 2013, Science, 342,
  598

\bibitem[{{Lyne} {et~al.}(2015){Lyne}, {Jordan}, {Graham-Smith}, {Espinoza},
  {Stappers}, \& {Weltevrede}}]{Lyne2015+Crab}
{Lyne}, A.~G., {Jordan}, C.~A., {Graham-Smith}, F., {et~al.} 2015, \mnras, 446,
  857

\bibitem[{{Marshall} {et~al.}(2016){Marshall}, {Guillemot}, {Harding},
  {Martin}, \& {Smith}}]{Marshall2016+B0540}
{Marshall}, F.~E., {Guillemot}, L., {Harding}, A.~K., {Martin}, P., \& {Smith},
  D.~A. 2016, \apjl, 827, L39

\bibitem[{{Mart{\'{\i}}n-Hern{\'a}ndez}
  {et~al.}(2003){Mart{\'{\i}}n-Hern{\'a}ndez}, {van der Hulst}, \&
  {Tielens}}]{MartinHernandez2003+HII}
{Mart{\'{\i}}n-Hern{\'a}ndez}, N.~L., {van der Hulst}, J.~M., \& {Tielens},
  A.~G.~G.~M. 2003, \aap, 407, 957

\bibitem[{{Menou} {et~al.}(2001){Menou}, {Perna}, \&
  {Hernquist}}]{Menou2001+Disk}
{Menou}, K., {Perna}, R., \& {Hernquist}, L. 2001, \apjl, 554, L63

\bibitem[{{Murphy} {et~al.}(2007){Murphy}, {Mauch}, {Green}, {Hunstead},
  {Piestrzynska}, {Kels}, \& {Sztajer}}]{Murphy2007+MGPS2}
{Murphy}, T., {Mauch}, T., {Green}, A., {et~al.} 2007, \mnras, 382, 382

\bibitem[{{Ostriker} \& {Gunn}(1969)}]{Ostriker1969}
{Ostriker}, J.~P., \& {Gunn}, J.~E. 1969, \apj, 157, 1395

\bibitem[{{Parent} {et~al.}(2011){Parent}, {Kerr}, {den Hartog}, {Baring},
  {DeCesar}, {Espinoza}, {Gotthelf}, {Harding}, {Johnston}, {Kaspi},
  {Livingstone}, {Romani}, {Stappers}, {Watters}, {Weltevrede}, {Abdo},
  {Burgay}, {Camilo}, {Craig}, {Freire}, {Giordano}, {Guillemot}, {Hobbs},
  {Keith}, {Kramer}, {Lyne}, {Manchester}, {Noutsos}, {Possenti}, \&
  {Smith}}]{Parent2011+J1119}
{Parent}, D., {Kerr}, M., {den Hartog}, P.~R., {et~al.} 2011, \apj, 743, 170

\bibitem[{{Pierbattista} {et~al.}(2015){Pierbattista}, {Harding}, {Grenier},
  {Johnson}, {Caraveo}, {Kerr}, \& {Gonthier}}]{Pierbattista2015+LCM}
{Pierbattista}, M., {Harding}, A.~K., {Grenier}, I.~A., {et~al.} 2015, \aap,
  575, A3

\bibitem[{{Pletsch} \& {Clark}(2014)}]{Methods2014}
{Pletsch}, H.~J., \& {Clark}, C.~J. 2014, \apj, 795, 75

\bibitem[{{Pletsch} \& {Clark}(2015)}]{Pletsch2015+J2339}
---. 2015, \apj, 807, 18

\bibitem[{{Pletsch} {et~al.}(2012){Pletsch}, {Guillemot}, {Allen},
  {et~al.}}]{Pletsch+2012-9pulsars}
{Pletsch}, H.~J., {Guillemot}, L., {Allen}, B., {et~al.} 2012, \apj, 744, 105

\bibitem[{{Ransom}(2001)}]{Ransom2001+Thesis}
{Ransom}, S.~M. 2001, PhD thesis, Harvard University

\bibitem[{{Ray} {et~al.}(2011){Ray}, {Kerr}, {Parent}, {et~al.}}]{Ray2011}
{Ray}, P.~S., {Kerr}, M., {Parent}, D., {et~al.} 2011, \apjs, 194, 17

\bibitem[{{Saz Parkinson} {et~al.}(2010){Saz Parkinson}, {Dormody}, {Ziegler},
  {Ray}, {et~al.}}]{SazParkinson2010+8BSPs}
{Saz Parkinson}, P.~M., {Dormody}, M., {Ziegler}, M., {Ray}, P.~S., {et~al.}
  2010, \apj, 725, 571

\bibitem[{Schwarz(1978)}]{Schwarz1978+BIC}
Schwarz, G. 1978, Ann. Stat., 6, 461

\bibitem[{{Stroh} \& {Falcone}(2013)}]{Swift+LATSurvey}
{Stroh}, M.~C., \& {Falcone}, A.~D. 2013, \apjs, 207, 28

\bibitem[{{Wakely} \& {Horan}(2008)}]{TeVCat}
{Wakely}, S.~P., \& {Horan}, D. 2008, International Cosmic Ray Conference, 3,
  1341

\end{thebibliography}
\end{document}